\documentclass{ws-ijgmmp}

\begin{document}

\newcommand{\be}{\begin{equation}}
\newcommand{\ee}{\end{equation}}
\def\bq{\begin{eqnarray}}
\def\eq{\end{eqnarray}}

\markboth{R. G. Vishwakarma}
{A Machian Approach to General Relativity}

\catchline{}{}{}{}{}

\title{A MACHIAN APPROACH TO GENERAL RELATIVITY}

\author{\footnotesize Ram Gopal Vishwakarma}

 \address{Unidad Acad$\acute{e}$mica de Matem$\acute{a}$ticas\\
 Universidad Aut$\acute{o}$noma de Zacatecas\\
 C.P. 98068, Zacatecas, ZAC, Mexico\\
Email: vishwa@uaz.edu.mx}

\maketitle

\begin{history}
\received{(Day Month Year)}
\revised{(Day Month Year)}
\end{history}

\begin{abstract}
Mach's principle is surely one of those tantalizingly beautiful concepts in physics which remain elusive.
Though General Relativity (GR) was conceived in the spirit of realizing it, the theory failed to fulfill this expectation.

Here a study on the implications of imposing Mach's principle on GR with an insight that spacetime has no  independent existence without a material background, is presented. This inclusion of the principle in GR turns out to be unexpectedly rewarding. The resulting theory solves many mysteries and averts lingering problems of the conventional forms of GR and cosmology.
\end{abstract}

\keywords{Gravitation; General Relativity - Fundamental Problems and General Formalism; Mach's Principle; Equivalence Principle and Electromagnetism.}

\ccode{PACS: 04.20.Cv, 04.20.-q, 95.30.Sf, 98.80.Jk}

\section{Introduction}

Einstein's masterwork - the theory of general relativity (GR) - a theory of unprecedented combination of mathematical elegance and conceptual depth, is recognized as an  intellectual achievement par excellence. 
The theory has made remarkable progress on both, theoretical and observational fronts \cite{progress} during a century since its inception (let us remember that the year 2015 marks the centennial of GR \cite{centennial}).
Nevertheless, it leaves a loose thread that scientists have struggled to tie up ever since: the theory fails to implement its own guiding principle - the Mach principle, which was considered earlier by Einstein as a philosophical pillar of GR. 

Though in the absence of a clear statement from Mach \cite{Mach}, there exist a number of formulations of Mach's principle \cite{Mach-many}, in essence it advocates eschewing the unobservable absolute space and time of Newton in favor of the directly observable background matter in the Universe. 
Mach critically examined the role of a background against which motion is to be measured and argued that, unless there is a material background, it is not possible to attach any meaning to the concepts of rest and motion.
That is, the acceleration of an object is determined not relative  to the absolute space but with respect to all other masses that occupy it. In other words, the inertia of the object depends on the background matter in the Universe, and
through the equivalence of inertia and gravitation, therefore, gravitational interaction would be impossible without the background matter. Hence, the principle implies that matter determines the curvature of spacetime.
Taken together with the principle of general covariance (non-existence of a privileged reference frame), which also emerges as a consequence of Mach's  denial of absolute space, this led to Einstein's field equations\footnote{As usual, $R_{ik}$ is the Ricci tensor defined by $R_{ik}=g^{hj} R_{hijk}$ in terms of the Riemann tensor $R_{hijk}$. $R=g^{ik}R_{ik}$ is the Ricci scalar, $g_{ik}$ is the metric tensor defined by $ds^2=g_{ik}dx^i dx^k$, $g^{ik}$ is the contravariant form of the metric tensor,  $T_{ik}$ is the energy-stress tensor of matter (which also includes in it any possible candidate of dark energy), $G$ is the Newtonian constant of gravitation and $c$ the speed of light in vacuum. The Latin indices assume values 0,1,2,3 unless stated otherwise.\medskip }
\be
R_{ik}-\frac{1}{2} g_{ik} R=-\frac{8\pi G}{c^4}T_{ik}.\label{eq:EinsteinEq}
\ee
Thus, Mach's far-reaching insight - that a physical theory should be based entirely on the directly observable quantities - played a leading role in Einstein's development of GR. Nevertheless, the theory failed to be perfectly Machian, contrary to Einstein's expectations.

Although some of Mach's ideas are indeed realized in the theory since GR asserts that the spacetime geometry is influenced by the matter distribution (rather than prescribing the spacetime geometry in advance), however there appear several anti-Machian features in GR. Firstly, there exists a class of curved solutions, which admit Einstein's equations  in the total absence of the energy-stress tensor, in which case Einstein's equations (\ref{eq:EinsteinEq}) reduce to
\be
R_{ik}=0.\label{eq:RicciEq} 
\ee
Not only some curved solutions of (\ref{eq:RicciEq}), but also its flat Minkowski solution possesses timelike geodesics and a well-defined notion of inertia, which render the theory non-Machian in the absence of any apparent conventional source.
Secondly,  there exists a class of cosmological solutions (for instance, the G$\ddot{o}$del solution \cite{Godel}) wherein the distant background rotates with respect to the local inertial frame defined by  gyroscopes, exhibiting a
preferred direction. This would be in conflict with Mach's principle which implies that a
local inertial frame should not rotate with respect to the frame defined by distant galaxies.

Although Einstein himself moved away from Mach's principle in his later years, the  principle continued to attract a lot of sympathy due to its aesthetic appeal, enormous impact and agreement with observation\footnote{Though some predictions of GR have been claimed to disagree with Machian expectations (see for instance \cite{Rindler}), however, this conclusion depends crucially on some particular formulation of Mach's principle used and a different formulation can lead to the opposite conclusion \cite{Bondi-Samuel}. Moreover, these claims should be reexamined in view of the new paradigm developed in the following sections.}. It is widely believed that  the principle must be implemented into any viable theory of gravitation.
As GR is the most successful theory of gravitation, one always aspires if GR could incorporate the tantalizingly lofty ideas of Mach. This would  be not just a wishful thought but an expectation warranted by the inner consistency and elegance of GR, and its consistency with special relativity (which too abolishes the absolute space) - all of which cry out with all force that GR must also be Machian.
The purpose of the present paper is to study the implications of imposing the ideas of Mach on the existing framework of GR.
Let us adopt a view, as Einstein had envisioned, that GR must be Machian.

\section{Quantifying Mach's Principle}\label{sec:DParameters}

To begin with, let us scrutinize the solutions of equations  (\ref{eq:RicciEq}) in the light of Mach's principle.
Among various statements of Mach's principle formulated in the context of GR, there is one version given by Pirani \cite{OS} that has been defined unambiguously: {\it ``In the absence of source, spacetime should necessarily be Minkowskian"}. 
 If the presence of source is defined by the presence of singularities (true gravitational ones where the Kretschmann scalar diverges) in the solutions of (\ref{eq:RicciEq}), as is in practice \cite{OS}, the principle would forbid a singularity-free solution from being anything but Minkowskian.
Hence the principle is violated in  the Ozsv$\acute{a}$th-Sch$\ddot{u}$cking solution \cite{OS}, the Taub-NUT (Taub-Newman-Unti-Tamburino) solution \cite{Taub} and a newly discovered solution \cite{Paper1} (given by equation (\ref{eq:int-kerr}) in the following), all of which are curved but singularity-free. 

It seems that the failure of the principle in these solutions warrants a deeper look and better understanding. Particularly, what does the curious presence of the timelike geodesics and a well-defined notion of inertia in these solutions  mean and why are they singled out in the absence of any conventional source? The presence of these trajectories of `particles under no force' must not be just coincidental and there must be some source.
Obviously we should expect some source of curvature in GR, wherein gravitation is a manifestation of the curvature of spacetime. 
This insinuates that something may be lacking in the theory, if it is correct. 
Perhaps we have to go to the very foundation of the formulation of source and define its presence in a more competent way which can apply universally to all  solutions with $R_{ik}=0$ but $R_{hijk} \neq 0$, and not through the singularity which is anyway an unphysical feature and a sign of breakdown of the theory.

\subsection{Source of curvature}\label{sec:KnownSols}

Mach's philosophy expects the source of curvature to be attributable entirely to some directly observable quantity, such as mass-energy, momentum, angular momentum or their densities. So, if GR is correct and it must be Machian, then
these quantities are expected to be supported by some parameters appearing in the curved spacetime solutions in such a way that the parameters vanish as the observable quantities vanish, reducing the solutions to the Minkowskian form. Further, these parameters (in the geometric units) are expected to be of the dimensions of the above-mentioned observable quantities (in the conventional units, $G$ and $c$ may also show up). Remarkably, it is indeed possible to write the curved solutions of equations  (\ref{eq:RicciEq}) in terms of such parameters, as we shall see in the following.

\bigskip
\noindent
{\sc Schwarzschild Solution:}
Although the source of curvature in the Schwarzschild solution
\be
ds^2=\left(1+\frac{K}{r}\right)c^2 dt^2-\frac{dr^2}{(1+\frac{K}{r})}-r^2d\theta^2-r^2\sin^2\theta ~d\phi^2,\label{eq:Sch}
\ee
 is assigned to the singularity present at $r=0$, it can also be defined alternatively through the parameter $K$ appearing as a constant of integration in the solution. The parameter appears in the Riemann tensor generatively and the tensor vanishes as $K$ vanishes. This ascertains that $K$ contains the source, which is also vouched by the well-known attribution of $K$ to an isotropic mass $M$ (placed at $r=0$) through $K=-2GM/c^2$
by requiring that GR should reduce to the Newtonian theory in the case of a weak, slowly-varying gravitational field.

\bigskip
\noindent
{\sc Kerr Solution:}
Similar possibility exists for the Kerr solution
\begin{eqnarray}\nonumber
ds^2 &=& \left(1+\frac{Kr}{r^2+\alpha^2 \cos^2\theta}\right)c^2 dt^2-\left(\frac{r^2+\alpha^2 \cos^2\theta}{r^2+Kr+\alpha^2}\right)dr^2 - (r^2+\alpha^2 \cos^2\theta)d\theta^2 \\\nonumber
&&-\left(r^2+\alpha^2-\frac{\alpha^2Kr}{r^2+\alpha^2 \cos^2\theta}\sin^2\theta\right)
\sin^2\theta ~d\phi^2\\
&&-\left(\frac{2\alpha Kr}{r^2+\alpha^2 \cos^2\theta}\right)\sin^2\theta~cdt~d\phi ,\label{eq:kerr}
\end{eqnarray}
which generalizes the Schwarzschild solution when the mass $M$ rotates as well. It is readily possible to assign the source of curvature in solution (\ref{eq:kerr}) to the source-carrier parameters $K$ and $\alpha$ without invoking the singularity. As is well known, the parameter $\alpha$ is represented as the angular momentum per unit mass: $\alpha=\kappa_1 J/(Mc)$, where $J$ is the angular momentum of the source mass and $\kappa_1$ indicates the possibility of a dimensionless constant.

\bigskip
\noindent
{\sc Taub-NUT Solution:}
As mentioned earlier, the Taub-NUT solution
\begin{eqnarray}\nonumber
ds^2 & =& \left(\frac{r^2+Kr-N^2}{r^2+N^2}\right)(c dt + 2N \cos \theta d\phi)^2 -\left(\frac{r^2+N^2}{r^2+Kr-N^2}\right) dr^2\\
&& -~(r^2+N^2)(d\theta^2+\sin^2\theta ~d\phi^2),\label{eq:Taub-NUT}
\end{eqnarray}
which is another generalization of Schwarzschild solution, does not possess any gravitational singularity. It is perfectly regular at $r = 0$ as its Kretschmann scalar does not blow up there. Though the solution possesses some spurious singularities, the so-called `wire singularities' at $\theta = 0$ and $\theta =\pi$ where the metric
fails to be invertible, these are only coordinate singularities which can be removed by introducing two coordinate patches \cite{Misner}.

Thus being curved but singularity-free, solution (\ref{eq:Taub-NUT}) cannot explain the source of its curvature in terms of the conventional source - singularity. Nevertheless, this situation can be averted and the solution can be made Machian if we define the source in terms of the `source carrier' parameters, which are $K$ and $N$ here. Clearly, the solution becomes Minkowskian for $K=N=0$. A simple dimensional analysis indicates that the parameter $N$ can be expressed in terms of the momentum $P$ of the source, i.e., 
$N=\kappa_2 GP/c^3$, where $\kappa_2$ is a possible dimensionless number.

The incompetence of the conventional representation of source through the singularities is also revealed through the following observation: When the Taub-NUT solution reduces to the Schwarzschild solution for a vanishing $N$, we assert the presence of the source  (the point mass $M$) at $r=0$. This source must not disappear (with the disappearance of singularity) when the mass gains additional features (expressed through $N$) in  (\ref{eq:Taub-NUT})! 
This strengthens our belief in the novel formulation of the source  in terms of the dimensional source-carrier parameters $K$ and $N$.

\bigskip
\noindent
{\sc Kerr-Taub-NUT solution:}
Competence of the novel strategy to represent the source through the source-carrier parameters is also demonstrated by the Kerr-Taub-NUT solution
\begin{eqnarray}\nonumber
ds^2 &=& \left(\frac{r^2+Kr+\alpha^2-N^2}{r^2+\{N+\alpha\cos\theta\}^2}\right) (cdt-\{\alpha\sin^2\theta-2N\cos\theta\}d\phi)^2\\\nonumber
&&-\left(\frac{\sin^2\theta}{r^2+\{N+\alpha\cos\theta\}^2}\right)  (\{r^2+\alpha^2+N^2\}d\phi-\alpha cdt)^2\\
&&-\left(\frac{r^2+\{N+\alpha\cos\theta\}^2}{r^2+Kr+\alpha^2-N^2} \right)dr^2
-(r^2+\{N+\alpha\cos\theta\}^2)d\theta^2,
\end{eqnarray}
which combines the Kerr and Taub-NUT solutions. The source of curvature can be efficiently attributed to the dimensional source-carrier parameters $K, \alpha, N$, in the absence thereof the solution reduces to the Minkowskian form.

\bigskip
\noindent
{\sc Kasner Solution:}
Though the Kasner solution in its standard form 
\be
ds^2=c^2 dT^2- T^{2p_1}dX^2- T^{2p_2}dY^2- T^{2p_3}dZ^2,\label{eq:kasnero}
\ee
does not contain any dimensional parameter which can be attributed to its source of curvature, it can easily be transformed to the form\footnote{The Kasner solution in this form was discovered by Narlikar and Karmarkar in 1946 \cite{curious}.}
\be
ds^2=c^2 dt^2- (1+nt)^{2p_1}dx^2- (1+nt)^{2p_2}dy^2
- (1+nt)^{2p_3}dz^2,\label{eq:kasner}
\ee
which serves the purpose. The dimensionless parameters $p_1$, $p_2$, $p_3$ appearing in (\ref{eq:kasnero}) and (\ref{eq:kasner}) satisfy
\be
p_1+p_2+p_3=1=p_1^2+p_2^2+p_3^2.\label{eq:kasnerIndices}
\ee
A simple dimensional analysis suggests that the parameter $n$, which has the dimension of the inverse of time in (\ref{eq:kasner}), can be expressed in terms of the momentum density, say ${\cal P}$, as $n=\kappa_3 \sqrt{G{\cal P}/c},~~ (\kappa_3={\rm a~possible ~dimensionless~number})$,
in order to meet its dimension.
As (\ref{eq:kasner}) reduces to the Minkowskian form for a vanishing ${\cal P}$,  the source of curvature present in (\ref{eq:kasner}), may be assigned to the net non-zero momentum density ${\cal P}$ arising from the expanding and contracting  spacetime\footnote{According to the standard interpretation, the Kasner solution represents space expanding and contracting at different rates in different directions. For instance, it expands in two directions and contracts in the third for $p_1=p_2=2/3$ and $p_3=-1/3$.}
given by (\ref{eq:kasner}). 

Although this interpretation does not make much sense in the conventional notion of empty space, nevertheless it is at least as good (or bad) as the standard interpretation of the solution in terms of the expanding/contracting space without any matter. If an utterly empty space without matter can expand and contract, it can very well  be associated with a momentum density. Moreover, the new interpretation provides useful clues to find new solutions of equations  (\ref{eq:RicciEq}), as explained in the following.

\bigskip
\noindent
{\sc A Solution Sourced by the Angular Momentum Density:}
This enlightening view of locating Mach's principle in the source of curvature through the source-carrier parameters, appears rewarding as  not only does it solve a lingering foundational problem, but also facilitates the discovery of  new solutions of equations  (\ref{eq:RicciEq}).  If a solution of equations  (\ref{eq:RicciEq}) can be supported by the momentum density as in the Kasner solution, we should also expect solutions of  (\ref{eq:RicciEq}) supported by the energy density and angular momentum density. Following these discerning guidelines, a new vacuum solution has been discovered recently \cite{Paper1} by defining a parameter $\ell$ in terms of the angular momentum density ${\cal J}$ as $\ell=G{\cal J}/c^3$. The solution reads\\
\be
ds^2 =\left(1-\frac{\ell^2x^2}{8}\right)c^2dt^2-dx^2-dy^2-\left(1+\frac{\ell^2x^2}{8}\right)dz^2
+\ell x (cdt-dz)dy+\frac{\ell^2x^2}{4}cdt~dz,\label{eq:int-kerr}
\ee
which is curved but singularity-free and hence provides the strongest support to the Machian strategy of representing the source in terms of the dimensional source-carrier parameters, here $\ell$.

Solution (\ref{eq:int-kerr}) as a new solution of the field equations  (\ref{eq:RicciEq}) is important in its own right. Moreover, it illuminates the so far obscure source of curvature in the Ozsv$\acute{a}$th-Sch$\ddot{u}$cking solution (\ref{eq:OS}), which would otherwise be in stark contrast with the new formulation of source, as it does not have any free parameter. In fact the Ozsv$\acute{a}$th-Sch$\ddot{u}$cking solution results from (\ref{eq:int-kerr}) by assigning a particular value to the parameter $\ell$: by defining new  coordinates $x_1, x_2, x_3, x_4$ through the transformation $x_1=y, \sqrt{2}x_2=-(ct+z),\sqrt{2} x_3=(ct-z), x_4=x$, it is easy to check that solution (\ref{eq:int-kerr}) for $\ell=2\sqrt{2}$ reduces to 
\be
ds^2=-(dx_1)^2 +4x_4 dx_1 dx_3 -2 dx_2 dx_3 -2 (x_4)^2 (dx_3)^2 - (dx_4)^2,\label{eq:OS}
\ee
which was discovered by Ozsv$\acute{a}$th and Sch$\ddot{u}$cking \cite{OS}. Another new solution of equations  (\ref{eq:RicciEq}) supported by the energy density is discovered in section \ref{sec:EngDen}.

\subsection{The new formulation of source is satisfying and insightful}

Thus all the curved solutions of equations (\ref{eq:RicciEq}) do contain dimensional parameters, which can support observables such as the energy, momentum or angular momentum or their densities. This means that without taking recourse to the singularity, the source of curvature in the absence of matter tensor can be defined more meaningfully by the ubiquitous presence of these parameters.  Since the parameters appear in the Riemann curvature tensor generatively, the tensor vanishes as the parameters vanish in the absence of the observable quantities, reducing the solutions to the Minkowskian form.
 It appears that if the presence of source in equations (\ref{eq:RicciEq}) is formulated in a modern way by the presence of such parameters, as should reasonably be expected, all solutions become Machian!

This enlightened way of defining the source of curvature in terms of the source-carrier parameters, presents a new scope to represent the source which applies universally to all solutions of equations (\ref{eq:RicciEq}), unlike the conventional source - singularity.
By approaching the very foundation of our ideas of spacetime, the new Machian strategy of defining the source appears more comprehensive and insightful than the conventional one  and grants the spacetime an existence with physical qualities of its own. This results in a {\it new paradigm} wherein the spacetime gets more and more to the status envisioned initially by Einstein.

\section{A New Insight}

Einstein revolutionized the notions of space and time and synthesized them into one spacetime. Yet, a still deeper vision of spacetime emerges in the new paradigm revealing that its novel Machian strategy  is subtler than may appear.

\subsection{Existence of spacetime implies existence of fields}

 What does the presence of the dimensional parameters sustaining observable quantities in the absence of matter tensor signify? It is obvious that the physical observable quantities - energy, momentum, angular momentum and their densities - have any meaning only in the presence of matter.  Thus the presence of such parameters in the solutions of equations (\ref{eq:RicciEq})  must not be just a big coincidence and  at face value, their ubiquitous presence in the solutions of (\ref{eq:RicciEq}) insinuates that fields are universally present in the spacetime in equations (\ref{eq:RicciEq}). 
That is, the presence of fields is implied by the presence of spacetime itself and this appears as a fundamental dictum of a Machian theory of gravitation.
As space is an abstraction from the totality of distance-relations between matter, it follows that the existence of matter (fields) is a necessary and sufficient condition for the existence of spacetime, and not only inertia, but also space and time should emerge from the interaction of matter.
Hence the spacetime is not something to which one can ascribe a separate
existence, independently of the matter fields. This is very much in the spirit of Mach's principle which implies that the existence of a spacetime structure has any meaning only in the presence of matter, which is bound so tightly to the spacetime that one can not exist without the other\footnote{Inspired by this, Einstein had envisioned that {\it ``space as opposed to `what fills space', has no separate existence"} \cite{Einstein} though he could not implement it in his field equations (\ref{eq:EinsteinEq}).}. 

This has however a far-reaching consequence. If the matter fields are expected to be present universally in the spacetime irrespective of its geometry, the flat Minkowskian spacetime should not be an exception and it must also be endowed with the matter fields and the ensuing gravitational field. [We shall see later (in section \ref{sec:flat}) how flatness is restored in the present formalism in spite of the presence of fields in the Minkowski spacetime.] 
However this renders the energy-stress tensor $T_{ik}$ as a redundant part of Einstein's equations (\ref{eq:EinsteinEq}), thereby ordaining the reduced equations (\ref{eq:RicciEq}) to serve as the field equations in the very presence of matter fields.

A deeper analysis of the equivalence principle also seems consistent with this new perspective.
As the equivalence principle renders gravitation as a geometric phenomenon,  the gravitational field (energy, momentum and angular momentum thereof) must also be revealed through the geometry itself and {\it not through the tensor $T_{ik}$} in (\ref{eq:EinsteinEq}). This would further imply, through the (local) equivalence of gravitation and inertia, that the inertial measures (energy, momentum and angular momentum) of matter should also be revealed through the geometry, through the metric field and {\it not through} $T_{ik}$.

Remarkably, this view fits very well with the nonexistence of a clean energy-stress tensor of matter. It may be mentioned that 
a recent study has discovered some surprising inconsistencies and paradoxes in the formulation of the energy-stress tensor of the  matter fields, concluding that the formulation of $T_{ik}$ does not seem consistent with the geometric description of gravitation \cite{vishwaApSS2}. 
Hence Mach's principle inherently requires the fields to appear through the spacetime itself, and not through the energy-stress tensor or singularity, which is usually taken for granted. 

In the classical tests of GR, we have anyway asserted the presence of the particulars of the source mass in the  Schwarzschild and Kerr solutions, {\it without invoking $T_{ik}$}. If this is so with two solutions of (\ref{eq:RicciEq}), so should it be with its other solutions. 
It is generally argued that equations (\ref{eq:RicciEq}) provide a meaningful representation only to space `outside' a massive body. However, we should note that the symmetry of a spacetime structure outside a mass in an otherwise empty space can also be shared by a spacetime structure inside a matter distribution, indicating that all the solutions of (\ref{eq:RicciEq}) may not necessarily represent an empty spacetime even in the standard paradigm.  

The picture thus emerges that the new insight - that spacetime has no separate existence from matter - 
takes GR to its logical extreme that the spacetime emerges from the interaction of matter. This
reconceptualizes the previous notion of spacetime by establishing it as the very source of gravitation.

\subsection{Geometry is governed by the net field}

As a matter field is always accompanied by the ensuing gravitational field (which also gravitates), and since  the matter is present universally in all spacetimes including the Minkowskian one, an important consequence of the adopted strategy is that {\it the geometry of the resulting spacetime is determined not by the matter fields alone, rather by the net contribution from the two fields.} This naive, self evident and plausible prediction that the metric field is entirely governed by the considered fields, is exactly what one would expect from a Machian theory.

Thus it is the resultant contribution from the material plus the gravitational fields, which determines the curved/flat status of a solution of the field  equations (\ref{eq:RicciEq}). If this resultant is non-zero, it appears as a dimensional parameter in the solution and the solution is curved. In the absence of a non-zero resultant, the solution becomes Minkowskian (unlike the conventional version of GR wherein this happens in the total absence of fields).
This is also corroborated by the study performed in section \ref{sec:DParameters}: as the parameters present in the curved solutions of equations (\ref{eq:RicciEq}) vanish, so does the Riemann tensor, reducing the solutions to the Minkowskian form.
This asserts that these parameters are the images of the  source of curvature.
Thus, a flat spacetime, which has so far been a notion of the ideal case of special relativity, can be achieved in the real Universe in the presence of matter/gravitation, which originates dynamically from the field equations, and is not just put by hand\footnote{In view of the vanishing of the Einstein tensor in a flat spacetime, the standard paradigm has to put  $T_{ik}$ equal to zero by hand in order to make equation (\ref{eq:EinsteinEq}) legitimate. Let us recall that the very derivation of $T_{ik}$ assumes its validity in the flat spacetime (special relativity).}.

We shall see in the following that this naive realization -  that the matter and the gravitational fields are present inherently in all spacetimes whose resultant determines geometry - naturally explains why there does not exist a proper energy-stress tensor of the gravitational field and why a homogeneous-isotropic Universe must be flat.

\subsection{Gravitational energy in the new paradigm}

As the gravitational field of a body can do work and must therefore contain energy, it must find a place in the `source term' 
 $T_{ik}$ in (\ref{eq:EinsteinEq}). Nevertheless, there is no place for the energy-stress of the gravitation field in $T_{ik}$, nor is there any way to define this energy-stress in a covariant way. A historical account of this issue would not be out of place. 

Einstein always viewed with suspicion the representation of the source of gravitation by the energy-stress tensor. First, he emphasized that this  `source term' should include all the sources of energy, momentum and stress, including those of the gravitational field. Although, the tensor $T_{ik}$ in (\ref{eq:EinsteinEq}) does include  in it all the sources of gravitation including the cosmological constant or any other dark energy candidate, but {\it except the gravitational field itself}. Failing to find a  tensor representation of the gravitational field, Einstein then admitted that {\it ``the energy tensor can be regarded only as a provisional means of representing matter''}. Alas, a century-long dedicated effort to discover a unanimous formulation of the energy-stress tensor of the gravitational field, has failed, concluding that a proper energy-stress tensor of the gravitational field does not exist.

According to the new insight however, this failure of the discovery of the energy-stress tensor of the gravitational field should not be surprising, as the energy, momentum and angular momentum of the gravitational field is already present in the spacetime geometry \cite{Vishwa_Front.Phys.}. In order to exemplify this, let us consider the Schwarzschild solution (\ref{eq:Sch}) in which matter (the point pass $M$) is present only at  $r=0$ whereas the solution is curved at all finite values of $r$. 
Hence the source of its curvature at all finite values of $r$ must be the gravitational energy, whose presence in the geometry is witnessed
 by the term $K/r=-2GM/(c^2r)$ present in (\ref{eq:Sch}). In view of the definition of the gravitational energy given by $-GM/r$ in the Newtonian theory (to which GR is supposed to reduce in the case of a weak slowly varying field), the term $-2GM/(c^2r)$ is just its relativistic analogue.

A compelling reason to assign the source of curvature in (\ref{eq:Sch}) to the gravitational energy, lies in the locality of GR. Let us note that locality becomes an intrinsic characteristic of GR as soon as the Newtonian concept of gravitation as a force (action-at-a-distance) is superseded by the curvature. Being a local theory, GR then assigns the curvature present at a particular point, to the source present at that very point. Thus, the agent responsible for the curvature  in (\ref{eq:Sch}) must be the gravitational energy, which is consistently defined at all  finite values of $r$ through the the term $K/r$, even though its source mass is present only at  $r=0$. 
Thus the long-sought-after gravitational energy of GR is already present in the geometry without taking recourse to the energy-stress tensor.

Similarly the angular momentum of the gravitational field, arising from the rotation of the mass $M$, is revealed through the geometry of the Kerr solution, as we have already seen.

\subsection{No empty space}

As has been mentioned earlier, the conventional belief is that only those curved solutions of equations (\ref{eq:RicciEq}) are meaningful which represent space outside some source matter in an otherwise empty space.  This however does not seem correct now. 
 The new paradigm nullifies a completely empty spacetime wherein equations (\ref{eq:RicciEq}) do not represent a spacetime completely devoid of fields\footnote{Thus the Schwarzschild and Kerr solutions in the new paradigm do not represent, in a strict sense, an empty spacetime outside a point mass; rather they represent an approximation of a local spacetime structure when the mass can be considered far away from the other matter in the Universe. \medskip}. 
Particularly, they do not always represent space outside a source mass. This is clear from the Kasner solution (\ref{eq:kasner}) and the new solution (\ref{eq:int-kerr}). A simple dimensional analysis reveals that the parameter $n$ in   solution (\ref{eq:kasner}) cannot be written in terms of the energy $E$, momentum $P$ or angular momentum $J$ such that the parameter vanishes when $E=P=J=0$.
The fact that the parameter $n$ can support only the density of momentum\footnote{The dimensional analysis can also suggest two other possible terms - densities of energy and angular momentum - in the expression for $n$. However, they vanish here: while the symmetries of (\ref{eq:kasner}) discard any possibility for the angular momentum density, the energy (density) disappears as it is canceled by the negative gravitational energy in a uniform matter distribution (as will also be ascertained in section \ref{sec:flat}). \medskip}, 
and {\it not} the momentum itself, asserts that solution  (\ref{eq:kasner}) results from a (uniform) matter distribution (throughout space) and not from a spacetime outside a point mass as in the cases of the Schwarzschild and Kerr solutions\footnote{This implies that the Kasner solution should emerge from a homogeneous distribution of matter expanding and contracting anisotropically (at different rates in different directions), which can give rise to a net non-zero momentum density represented through the parameter $n$ serving as  the source of curvature.
  Thus the new paradigm provides a meaningful explanation to the Kasner solution unlike the standard paradigm wherein the  physical meaning of the Kasner solution remains an unsolved mystery in the context of the conventional empty spacetime.}.

The same is true about the new solution (\ref{eq:int-kerr}), wherein the parameter $\ell$ can support only the density of angular momentum, and {\it not} the angular momentum itself. This asserts  that solution (\ref{eq:int-kerr}) results from a rotating {\it matter distribution} (confined to
$-\frac{2\sqrt{2}}{|\ell|}<x<\frac{2\sqrt{2}}{|\ell|}$) and not from a spacetime outside a point mass. 

We thus see that various evidences of the presence of fields do exist in the solutions of the field equations (\ref{eq:RicciEq}). 
A natural question then arises that if the new paradigm, governed by the field equations (\ref{eq:RicciEq}), asserts the presence of fields universally, why and how do some of its solutions resume curvature while others flatness? This question is not answered satisfactorily in the standard paradigm, wherein the notion of the presence/absence of fields in (\ref{eq:RicciEq}) is not defined properly. 
The facts, that an empty spacetime devoid of any field is inexistent in the new paradigm,  and the geometry of a spacetime is determined by the net field, are crucial for rendering the homogeneous-isotropic spacetime flat. We shall note this in the following.

\subsection{A homogeneous-isotropic Universe becomes Minkowskian}\label{sec:flat}

It would be curious to note that a homogeneous, isotropic solution of equations (\ref{eq:RicciEq}) becomes flat. This solution can be obtained by solving equations (\ref{eq:RicciEq}) for the Robertson-Walker metric, giving
\be
ds^2=c^2 dt^2-c^2t^2\left(\frac{dr^2}{1+r^2}+r^2d\theta^2+r^2\sin^2\theta ~d\phi^2\right),\label{eq:milne}
\ee
which represents a homogeneous space expanding (or contracting) isotropically. The solution can be reduced to the Minkowskian form by using the transformations $\bar{t}=t\sqrt{1+r^2}$, $\bar{r}=ctr$ \cite{narlikar}.

It would prove challenging for the standard paradigm to explain why solution (\ref{eq:milne}) is flat, while the rest of the solutions of equations (\ref{eq:RicciEq}), viz., the Schwarzschild, Kerr, Kasner, Taub-NUT, Kerr-NUT and the new solution (\ref{eq:int-kerr}), all  are curved.
Particularly, how a simple change from anisotropy to isotropy can reduce the curved Kasner solution (\ref{eq:kasner}) into a flat solution (\ref{eq:milne}), cannot be answered by the conventional wisdom.

Remarkably, a convincing resolution comes from the new paradigm only.
It may be noted that an isotropic expansion or contraction ($p_1=p_2=p_3\neq 0$) is not possible in the Kasner spacetime  (\ref{eq:kasner}) in general, as $p_1$, $p_2$, $p_3$ are constrained by (\ref{eq:kasnerIndices}). Nevertheless, this becomes possible when $n=0$, in which case solution (\ref{eq:kasner}) reduces, independently of (\ref{eq:kasnerIndices}), to solution (\ref{eq:milne}) (written in different coordinates). This  insinuates that if a homogeneously distributed matter expands or contracts isotropically,  its net momentum density vanishes and the resulting spacetime becomes flat. In other words, the energy density of a homogeneously distributed matter does not contribute to the curvature of spacetime. The only way this can happen is that the positive energy of a uniformly distributed matter be canceled precisely by the negative gravitational energy at each point. Thus the new paradigm predicts that {\it the sum of the matter and gravitational energies must be vanishing in the Universe of a uniform matter distribution}.

This appears consistent with the other solutions of equations (\ref{eq:RicciEq}) in the framework of the new paradigm, which guarantees the presence of the material fields and the ensuing gravitational field in the field equations (\ref{eq:RicciEq}).
According to this, solution (\ref{eq:milne}) would then represent a homogeneously distributed matter throughout the space at all times, which is either expanding or contracting isotropically.
As the positive energy of the matter field would be exactly balanced, point by point,  by the negative energy of the resulting gravitational field  (contrary to the case of the Schwarzschild solution where there would be only the gravitational energy and no matter at the points represented by the solution), this would provide a net vanishing energy. Neither would there be any net momentum contribution from the isotropic expansion or contraction of the material system (contrary to the case of the Kasner solution).  Hence, in the absence of any net non-zero energy, momentum or angular momentum, solution (\ref{eq:milne}) would not have any curvature.

This prediction may not appear too dubious if we take full account of gravitation by considering not only its gravito-electric aspect (as in the Newtonian gravity) but also the gravito-magnetic aspect, which does appear in GR \cite{gravitoEM, Bahram1}.
Interestingly Sciama \cite{Sciama} has already shown, by using the gravito-electric and gravito-magnetic aspects of gravity in his Machian theory of inertia (wherein the inertia of an object is mathematically equal to the reaction force produced by the combined gravity of all matter in the Universe upon the object), that the total energy, gravitational plus inertial, of a particle at rest in the Universe is zero.
This also appears consistent with several other investigations and results which indicate that the total energy of the Universe is zero. Hawking and Milodinow \cite{Hawking-Milodinow} have argued recently that the total energy of the Universe must always remain zero, as the positive energy of the matter can balance the negative gravitational energy.

This view would also be consistent with a broader picture of the equivalence principle:
As a uniform gravitational field can be abolished  everywhere in a suitable inertial (freely falling) reference frame, the gravitational energy at every point inside a uniformly distributed matter must also be canceled by the inertial energy present at that point.  

The results obtained in this and the preceding sections, taken together with the above-mentioned result from Sciama, provide an important insight that the formulation of the gravitational energy should be sought in the gravito-electromagnetic aspect of GR (and not in the energy-stress tensor which would be ruled out in a Machain theory, as we have seen).

It is now well-established that the gravito-electromagnetism is an inherent aspect of GR.
Unlike the Newtonian gravity, a current of mass-energy generates a gravitomagnetic field in GR.
Maartens and Bassett  \cite{gravitoEM} have shown that the free gravitational field possesses a close analogy between the Maxwell equations for the electromagnetic fields and the Bianchi identities for the gravito-electromagnetic fields.
It has also been shown that in such fields, the axes of gyroscopes exactly follow the predictions made by Mach's principle \cite{Schmid}.
Remarkably, this result remains valid in a Universe, which is arbitrarily close to the Milne model.
As has been mentioned earlier, some predictions of gravitomagnetism are claimed to disagree with Machian expectations \cite{Rindler}. However, this conclusion depends on some particular formulation of Mach's principle used and a different formulation changes the conclusion \cite{Bondi-Samuel}. 

Various attempts, performed so far and ongoing, to test the gravitomagnetism (under the denomination of `frame-dragging' - dragging of inertial frames by rotating bodies) with natural and artificial objects, are reviewed in \cite{testGravitoMag}.

\subsection{Status of the new theory on Machianity}

The novel feature GR would now acquire - that the spacetime solutions of equations (\ref{eq:RicciEq}), including the Minkowskian one,  are not devoid of fields - is remarkably crucial for the theory on Machianity.
The ubiquitous presence of fields not only renders the resulting theory consistent with Mach's principle, but also explains the curious existence of the timelike geodesics and the well-defined notion of inertia in the curved as well as flat solutions of equations (\ref{eq:RicciEq}).
This also provides a meaningful explanation to various so-far unexplained solutions, such as the Kasner solution, the Ozsv$\acute{a}$th-Sch$\ddot{u}$cking solution, etc., which are generally regarded unphysical. 

As has been mentioned in the beginning, there exists a second category of the non-Machian solutions in the standard paradigm, for instance the  G$\ddot{\rm o}$del solution, which exhibits an intrinsic rotation of matter with respect to the local inertial system and admits closed timelike curves\footnote{The G$\ddot{o}$del-type solutions, which admit closed timelike curves and hence permit possibility to travel in the past, violate the concepts of causality and  create paradoxes: ``what happens if you go back in the past and kill your father when he was a baby!'' Rightly this kind of solutions is absent from the new paradigm.}. Remarkably, these are absent in the new paradigm due to the absence of $T_{ik}$, which strengthens our confidence in the Machian strategy we adopted to quantify Mach's principle.

\subsection{Naive field equations with desirable qualities}

We have seen that the new insight - that the matter fields, together with the ensuing gravitational field, are already present inherently in the spacetime geometry - renders $T_{ik}$ a redundant part of Einstein's field equations (\ref{eq:EinsteinEq}), and hence establishes the elegant and deceptively simple  field  equations (\ref{eq:RicciEq}) as the new field equations. 

The new field equations are vouched by the simplicity of its Lagrangian density - the scalar curvature $R$ - which is the simplest invariant of a geometrical origin.
It is well-known that by considering the scalar curvature as the Lagrangian density for the Einstein-Hilbert action, the Euler-Lagrange equations under the variations in the metric, constitute the field equations (\ref{eq:RicciEq}), which govern the gravitational interactions of matter appearing now geometrically. This interaction is governed by the Bianchi identities (appearing as as the propagation equations), now given in terms of the Weyl tensor - the trace-free part of the Riemann curvature tensor.

Interestingly, the resulting theory becomes scale invariant (as any physical theory is desired to be) since the new field equations (\ref{eq:RicciEq}), are manifestly scale invariant. It may be noted that the complete equation (\ref{eq:EinsteinEq}), unlike the rest of physics, is not scale invariant, which is regarded as a major defect of GR.

We have already seen that the new theory based on the field equations (\ref{eq:RicciEq}) is free from all the non-Machian features usually encountered in the standard paradigm.

\section{\bf Internal Consistency of the New Paradigm}

The new paradigm receives strong support from the observations at all scales. The classical tests of GR, which have  
successfully tested the relativistic gravity effects in  weak gravity conditions, have in fact substantiated the new paradigm in the solar system and binary neutron stars. Interestingly, the new paradigm, in the guise of the Milne model, successfully explains the cosmological observations as well, without requiring the dark sectors of the standard paradigm.

\subsection{A simple, elegant Universe with long-sought-after properties}

In order to explain the present cosmological observations, the standard cosmology requires two mysterious, invisible, and as yet unidentified ingredients - dark matter and dark energy -  neither of which is satisfactorily understood.
One the one hand, the theory predicts that about $27 \%$ of the total content of the Universe is made of non-baryonic dark matter particles, which should certainly be predicted by some extension of the Standard Model of particles physics. However, there is no indication of any new physics beyond the Standard Model of particle physics which has been successfully verified at the Large Hadron Collider. Rightly, the dark matter has eluded our every effort to directly observe it.
On the other hand, the dark energy, which is believed to constitute about $68 \%$ of the total content of the Universe, is even more difficult to explain than dark matter, besides posing other theoretical challenges. 

In this context, it would be remarkable to note that all the cosmological observations can be explained successfully in the framework of Milne's model,  without requiring the dark components of the standard cosmology  \cite{Vishwa_Front.Phys., vishwa_milne}. Moreover, the model averts in a natural way the long-standing problems of the standard cosmology, such as the horizon, flatness and the cosmological constant problems, as has been shown in \cite{Vishwa_Front.Phys., vishwa_milne}. This may seem like an extraordinary coincidence, as the Milne model is not believed to represent the real Universe!

Milne model, based on the kinematic relativity and cosmological principle, was in fact developed independently of GR, though it is mathematically equivalent to the homogeneous, isotropic solution  (\ref{eq:milne}) of equations (\ref{eq:RicciEq}). The model assumes the presence of matter in the Minkowskian background and serves as a phenomenological model of the Universe  \cite{vishwa_milne, Bondi}. Thus it is not an unphysical (empty) model, as is generally (mis)believed. Nevertheless, in the absence of a concrete theory it remained an ad-hoc theory and could not explain why the presence of matter should not create a possible curvature in the spacetime.

Remarkably,  the new paradigm unravels the mystery how a Universe with homogeneous, isotropic distribution of matter naturally becomes Minkowskian, which is nothing but the Milne model.  Thus the new paradigm, which naturally leads to the Milne model in a cosmological scenario, provides a rigorous theoretical foundation to the Milne model.
Hence the mysterious coincidences of the successful explanation of the cosmological observations in terms of the Milne model, provide a powerful evidence of the correctness of the new paradigm. The observations actually reveal a simpler and more elegant Universe than anyone could have imagined! 

We have noted that the resulting cosmology contains two time scales which are very useful. The line element (\ref{eq:milne}) uses the comoving coordinates and a cosmic time, in terms of which the Universe appears dynamic. This provides a suitable reference frame to explain the cosmological observations.  However, solution (\ref{eq:milne}) can also be transformed to the Minkowskian form in the locally defined measures of space and time, in terms of which the big bang singularity\footnote{Although the spacetime given by the line element (\ref{eq:milne}) does not have any singularity at $t=0$, however the volume of its spatial sections vanishes at $t=0$. This causes the density of the accompanied matter to diverge, creating a big bang-like situation.} 
is circumvented.
As all the candidates of dark energy are incorporated in $T_{ik}$, its disappearance from the theory circumvents the dark energy crisis in a natural way. 

Remarkably, the symmetries of the Minkowski space make it possible to validate the conservation of the energy in the Universe in the new paradigm, solving the long-standing problems associated with the conservation of energy. It has been shown by  Noether that the symmetry of the Minkowskian space is the cause of the conservation of the energy-momentum of a physical field which otherwise leads inevitably to deep difficulties with the definition and conservation in a curved background \cite{Noether}.

\subsection{Classical limit of $R_{ik}=0$ in the new paradigm}

Highly successful in everyday applications, the Newtonian theory of gravitation also provides excellent approximations under a wide range of astrophysical cases \cite{NewtonianGrav}. Hence the first crucial test of a relativistic theory of gravitation is that it reduces to the Newtonian gravity for a slowly varying weak gravitational field where the concerned velocities are much less than $c$. However, the standard paradigm fails to fulfill this requirement as equations (\ref{eq:EinsteinEq}), in the limit of the weak field, do not reduce to the Poisson equation in the presence of a non-zero cosmological constant $\Lambda$ (or any other candidate of dark energy), which becomes unavoidable in the standard paradigm. It  is though dropped from equations (\ref{eq:EinsteinEq}) in local problems,  by assuming that  $\Lambda\approx 10^{-56}$ cm$^{-2}$ (as inferred from the cosmological observations) must not contribute to the physics appreciably there. Nevertheless, it has been shown recently that even this value of $\Lambda$ does indeed contribute to the bending of light and to the advance of the perihelion of planets \cite{Ishak}.

In this context, it would be encouraging to note that equations (\ref{eq:RicciEq}), in the new paradigm, consistently admit the Poisson equation in the limit of a weak gravitational field, provided we take into account the inertia of matter besides its gravitational aspects (as should be expected in a true Machian theory).

Let us consider a spacetime structure resulting from a matter distribution of low proper density moving with low velocity. In this case the spacetime can be approximated by a time-independent line element given by the Lorentz metric plus a small time-independent perturbation $h_{ik}$:
\be
g_{ik}=\eta_{ik}+h_{ik}.\label{eq:L+pert}
\ee
It is well-known that for (\ref{eq:L+pert}), the geodesic equation in the first order of approximation, reduces to
\be
\frac{d^2\vec{ r}}{dt^2}=-\frac{c^2}{2}\vec{\nabla} h_{tt}\label{eq:geoEq}
\ee
and the Ricci tensor $R_{ik}$ reduces to $\nabla^2 h_{tt}/2$, so that the field equations (\ref{eq:RicciEq}) reduce in the limit of weak fields to
\be
\nabla^2 h_{tt}=0,\label{eq:htt}
\ee
where $h_{tt}$ is the time-time component of $h_{ik}$.
We have already witnessed in various examples the ubiquitous presence of the gravitational and the inertial (material) fields in the geometry. 
In the Newtonian picture, these fields are to be recast in terms of forces, i.e., the inertia of matter and the gravitation of matter, whose resultant is expected to drive the motion of a test particle through Newton's second law of motion
\be
\frac{d^2\vec{ r}}{dt^2}=-\vec{\nabla}\phi_{\rm net}=-\vec{\nabla}(\phi_{\rm g}+ \phi_{\rm i}),\label{eq:eqMotion}
\ee
where $\phi_{\rm g}$ and $ \phi_{\rm i}$ are respectively the gravitational and inertial potentials whence are derived the corresponding forces\footnote{It should be noted that the two potentials $\phi_{\rm g}$ and $ \phi_{\rm i}$ characterize two different interactions: while the gravitational potential is measured, in the limit of the weak field, in terms of the work done by bringing unit mass from infinity to the point in question, the inertial potential, according to Mach's principle, is determined from an interaction between the mass and the rest of the matter in the Universe.\medskip}. 
The relativistic equation of motion (\ref{eq:geoEq}) is seen to be precisely the Newtonian equation of motion (\ref{eq:eqMotion}) if we identify $h_{tt}$ by 
\be
 h_{tt} = \frac{2}{c^2}(\phi_{\rm g}+ \phi_{\rm i}).\label{eq:G&I}
\ee
Equation (\ref{eq:htt}) then reduces to
\be
\nabla^2 \phi_{\rm g}=- \nabla^2 \phi_{\rm i}.\label{eq:PoissonT}
\ee
Equation (\ref{eq:PoissonT}) just sets a simple constraint on how the two fields are related, otherwise leaving all the possibilities open for the details of the individual fields. Thus the inevitable presence of matter in the new paradigm, taken together with the reduction of the geodesic equation to the Newtonian equation of motion, indicates that the gravitational potential $\phi_{\rm g}$ should follow the precepts of the Newtonian gravitation in the limit of the weak field. In other words, there is nothing in the present case to contradict the Poisson equation\footnote{The same is not true in the standard paradigm wherein the field equations (\ref{eq:EinsteinEq}) reduce, in the limit of a weak field, to a Poisson-like equation $\nabla^2 \phi_{\rm g}=4 \pi G(\rho+\rho_{de})$ (with $\rho_{de}$ denoting the density of the dark energy), which is not consistent with (\ref{eq:Poisson}) unless $\rho_{de}$ is vanishing.}
\be
\nabla^2 \phi_{\rm g}=4 \pi G\rho_{\rm g}\label{eq:Poisson}
\ee
for the field $\phi_{\rm g}$, where $\rho_{\rm g}$ is the gravitational mass density of the considered matter distribution.
This, taken together with (\ref{eq:PoissonT}), implies that the inertial field $\phi_{\rm i}$ too holds a Poisson-like equation in the limit of a weak field, as is rightly expected in a Machian theory. This indicates that the inertia of matter is an inductive reaction force produced by the combined gravitational effect of all matter in the Universe, as has also been shown by Sciama by using the gravito-electromagnetic aspects of gravitation in his Machian theory of inertia \cite{Sciama}.

\subsection{Why the speed of light is so unique}

We have noted that the sum of the gravitational and inertial energies from a uniform matter distribution should be vanishing, which explains the flatness of the Milne model. We shall see that it also provides a new insight on why the speed of light is so special - why it is same in all reference frames. This appears so, just because the speed of light is determined by the total gravitational potential of the Universe.

 Let us consider a uniform matter distribution of radial size ${\cal R}$ and mass ${\cal M}$, and gradually increase its size ${\cal R}$ (hence mass ${\cal M}$) close to that of the (observable) Universe. Hence the gravitational (potential) energy at a point just outside the chosen matter distribution can be approximated in terms of the Newtonian gravitational potential, giving the inertial/gravitational energy per unit mass as
\be
{\rm Inertial ~energy}=|{\rm Gravitational ~energy}|=\kappa \frac{G{\cal M}}{{\cal R}},\label{eq:IE}
\ee
where $\kappa$ is a constant of order unity. This formulation of the gravitational potential is not way off base as it consistently reproduces the  Schwarzschild case for $\kappa=2$. (Let us also note that an isotropic mass distribution appears gravitationally to an external observer as though all of its mass were concentrated at its centre.)
Since an estimate of this inertial energy, according to $E=mc^2$, would be $c^2$,  (\ref{eq:IE}) would imply 
\be
c=\sqrt{\frac{\kappa G{\cal M}}{{\cal R}}}.\label{eq:c}
\ee
In the limiting case of ${\cal R}$ and ${\cal M}$ as respectively approaching the size and the mass of the Universe, equation (\ref{eq:c}) provides an important new insight that the speed of light is determined by the total inertial (or gravitational) potential of the Universe! This explains why a mass multiplied by $c^2$ gives the energy stored in it!
Equivalently, (\ref{eq:c}) can give $G$ in terms of the total gravitational potential of the Universe. 
Interestingly the relation  (\ref{eq:c}) has also been derived alternatively by Sciama from the gravito-electromagnetic  aspects of his Machian theory of inertia \cite{Sciama}.
This interconnection of $c$ and $G$ to the distribution of matter in the Universe, is missing in the conventional GR.

\subsection{A real Universe without $T_{ik}$}

A constrained mind, used to visualize matter through $T_{ik}$ only, would find it difficult to apprehend the real Universe composed of planets, stars and galaxies in terms of equations  (\ref{eq:RicciEq}). But, do we not explain successfully various gravitational phenomena in the solar system (experienced through various tests of GR) in terms of the  Schwarzschild and Kerr solutions of equations  (\ref{eq:RicciEq})?  The presence of matter in the new paradigm may seem less audacious than it is in the standard paradigm and only the resultant of the inertial (matter) and gravitational fields is revealed through the dimensional parameters present in the spacetime metric. Nevertheless, the presence of matter can also be witnessed in terms of the conventional quantities. For example, if we estimate ${\cal R}$ appearing in equation (\ref{eq:c}) in terms of the scale factor of a dynamical form of the Universe given by (\ref{eq:milne}), we can write it as
\be
{\cal R}=cH^{-1},
\ee
where the Hubble parameter $H=1/t$ for the model  (\ref{eq:milne}).  Equation (\ref{eq:c}) then reduces to
\be
H^2=\frac{4 \pi \kappa G}{3}\rho,
\ee
where $\rho$ is the density of the homogeneously distributed matter at the epoch $t$. This is but the Friedmann equation of the standard cosmology for $\kappa=2$, which though appears here without invoking $T_{ik}$!

As has been mentioned above, the symmetries of the Minkowski space (which represents homogeneous, isotropic Universe in the new paradigm) make it possible to validate the conservation of energy in the Universe. This means that the continuity and the Navier-Stokes equations of fluid dynamics would hold in the new paradigm. This may provide a framework to understand the structure formation in the Universe in terms of a process involving the interplay of the gravitational interaction and the pressure force, as was considered by Jeans in the Newtonian Cosmology \cite{narlikar}. The same was perhaps envisioned by Milne though could not be completed \cite{Bondi}. The new paradigm thus  provides a promising hope to complete Milne's unfinished task.

\section{\bf A Solution Sourced by the Energy Density - A New Solution of $R_{ik}=0$}\label{sec:EngDen}

As has been mentioned earlier, the new paradigm predicts the existence of a solution of equations (\ref{eq:RicciEq}) whose curvature can be supported by the energy density. To complete the theory, we try to discover this solution here. 
 
 As we have noted earlier, the resultant of the matter and gravitational energy densities vanishes in a homogeneous spacetime, since the positive matter energy is cancelled by the negative gravitational energy at each point. Hence, this resultant, in an  inhomogeneous spacetime, is expected to be non-vanishing which will stand out as a contribution to the source of curvature of such spacetime. Therefore a solution of equations (\ref{eq:RicciEq}) whose curvature can be supported only by the energy density, is expected, in the simplest case, to be an inhomogeneous, static and non-rotating line element so that its curvature does not receive contributions from the momentum and angular momentum densities.

Such a line element can be obtained by generalizing/inhomogenizing the Minkowski metric with the help of a parameter which can sustain the energy density. 
With the help of the energy density ${\cal E}$, $G$ and $c$, one can construct the following two parameters (whose dimensions can be written in terms of the coordinates, i.e., space or time), so that the parameter vanishes when ${\cal E}$ vanishes:
\begin{enumerate}
\item
~~~$\mu=\frac{G{\cal E}}{c^4}$, 

\item
~~~$\nu=\frac{G{\cal E}}{c^2}$, 
\end{enumerate}
which have the dimensions of the inverse of the square of length and the inverse of the square of time respectively. Clearly, the first choice is suitable for a static spacetime. 
As the required solution, expected to be inhomogeneous, should be sourced by $\mu$ so that it reduces to the Minkowskian form for $\mu$ = ${\cal E}$ = 0, let us generalize/inhomogenize the Minkowski metric with the help of the factors $(1+a_i \mu X_i^2)^{\gamma_i}$, where $X_i$ is a space coordinate of the dimension of length appearing in the considered form of the Minkowski metric and $a_i, \gamma_i$ are some dimensionless numbers.

For instance, by considering the Minkowski metric in the spherical polar coordinates, wherein $r$ is the only coordinate of the dimension of length, let us consider the following line element
\begin{eqnarray}\nonumber
ds^2&=&(1+a_0 \mu r^2)^{\gamma_0}c^2 dt^2-(1+a_1 \mu r^2)^{\gamma_1} dr^2-(1+a_2 \mu r^2)^{\gamma_2}r^2d\theta^2\\
&&-(1+a_3 \mu r^2)^{\gamma_3}r^2\sin^2\theta ~d\phi^2
\end{eqnarray}
as a possible solution of equations  (\ref{eq:RicciEq}) for some values of $a_i, \gamma_i$.  However, no non-trivial solution is obtained in this case. A similar exercise with 
the Minkowski metric considered in the Cartesian coordinate system (which has three coordinates  of the dimension of length: $x, y, z$), makes the solution intractable. A breakthrough ensues when the Minkowski metric is considered in the cylindrical coordinates, which have two coordinates of the dimension of length: $r$ and $z$. That is, we consider the line element
\begin{eqnarray}\nonumber
ds^2&=&(1+a_0 \mu r^2)^{\gamma_0} (1+b_0 \mu z^2)^{\lambda_0} c^2 dt^2-(1+a_1 \mu r^2)^{\gamma_1} (1+b_1 \mu z^2)^{\lambda_1} dr^2\\\nonumber
&&-(1+a_2 \mu r^2)^{\gamma_2}(1+b_2 \mu z^2)^{\lambda_2}  r^2 d\phi^2\\
&&-(1+a_3 \mu r^2)^{\gamma_3} (1+b_3 \mu z^2)^{\lambda_3}dz^2.\label{eq:ED}
\end{eqnarray}
The constants $a_i, b_i, \gamma_i, \lambda_i$ appearing here are determined by solving (\ref{eq:RicciEq}) for (\ref{eq:ED}), giving 
\begin{eqnarray}\nonumber
& \gamma_0= \gamma_3=-2,  ~~~ \gamma_1=-4,  ~~~ \gamma_2=0,\\\nonumber
&\lambda_0=1, ~~~ \lambda_1= \lambda_2=0, ~~~ \lambda_3=-1,\\\nonumber
& a_0 = a_1=a_3=1, ~~~ b_0=b_3=4,
\end{eqnarray}
so that (\ref{eq:ED}) reduces to the following simple form
\be
ds^2=\frac{(1+4\mu z^2)}{(1+ \mu r^2)^2}c^2 dt^2 - \frac{ dr^2}{(1+ \mu r^2)^4} - r^2 d\phi^2 - \frac{dz^2}{(1+4 \mu z^2)(1+ \mu r^2)^2}.\label{eq:EneDen}
\ee
This is the expected line element and appears as another new, exact and curved solution of equations (\ref{eq:RicciEq}), which results from a inhomogeneous axisymmetric distribution of matter, according to the new paradigm. 

As solution  (\ref{eq:EneDen}) is curved but singularity-free for all finite values of the coordinates [this is also corroborated by its Kretschmann scalar = $192 \mu^2 (1+\mu r^2)^6$, which is well-behaved for all finite values of $r$], it provides, in the absence of any conventional source there, a powerful support to the Machian strategy of representing the source in terms of the dimensional source-carrier parameters, here $\mu$.
We note that $\mu$, and hence ${\cal E}$, do source the curvature of solution (\ref{eq:EneDen}) which becomes flat for $\mu=0$, in consistence with our initial expectations. 

Solution (\ref{eq:EneDen}) indeed appears new, as no well-known solutions of equations (\ref{eq:RicciEq}) (already listed in section \ref{sec:KnownSols}) can support the density of energy in the case of a static, non-rotating spacetime, as we have already seen. 
The non-vanishing components of the Christoffel symbol and the Riemann tensor, for solution (\ref{eq:EneDen}), are obtained as the following.

\begin{equation}
 \left.\begin{aligned}
 \Gamma^t_{tz}=-\Gamma^z_{zz}=\frac{4\mu  z}{1+4\mu z^2},~~~ \Gamma^t_{tr}=\frac{\Gamma^r_{rr}}{2}=\Gamma^z_{rz}=-\frac{2\mu r}{1+\mu r^2},\\
\Gamma^r_{tt}=-2\mu r(1+\mu r^2)(1+4\mu z^2), ~~~ \Gamma^r_{\phi \phi}=-r(1+\mu r^2)^4,\\
  \Gamma^r_{zz}=\frac{2\mu r(1+\mu r^2)}{1+4\mu z^2}, ~~~ \Gamma^\phi_{r \phi}=\frac{1}{r}, ~~~ \Gamma^z_{tt}=4\mu z(1+4\mu z^2) ;
       \end{aligned}
 \right\}
\end{equation}

\medskip
\begin{equation}
 \left.\begin{aligned}
 R_{trtr}=\frac{2\mu(1+4\mu z^2)}{(1+\mu r^2)^3}, ~~~ R_{t\phi t\phi}=2\mu r^2(1+\mu r^2)(1+4\mu z^2),\\
 R_{tztz}=-\frac{4\mu}{1+\mu r^2}, ~~~ R_{r\phi r\phi}=\frac{4\mu r^2}{1+\mu r^2},\\
 R_{rzrz}=-\frac{2\mu}{(1+\mu r^2)^3(1+4\mu z^2)}, ~~~ R_{\phi z\phi z}=-\frac{2\mu r^2(1+\mu r^2)}{1+4\mu z^2}.
       \end{aligned}
 \right\}
\end{equation}

\section{Interior Solutions}

The usual Schwarzschild interior solution, providing the standard representation of the interior of a static spherically symmetric non-rotating star, assumes  a static sphere of matter consisting of an incompressible perfect fluid of a constant density. This however turns out to be unphysical, since the speed of sound $=c\sqrt{dp/d\rho}$ becomes infinite  in the fluid with a constant density $\rho$ and a variable pressure $p$. It appears that the conventional Schwarzschild-interiors are not conceptually satisfying.
The Kerr solution has remained unmatched to any known 
non-vacuum solution that could represent the interior of a rotating star. It seems that we have been searching for the interior solutions in a wrong place. 

As the new paradigm claims that equations (\ref{eq:RicciEq}) provide a realistic theory of gravitation in the presence of matter, the interior solutions should also be searched therein.
We shall see in the following that the predictions of the new paradigm, in connection with the interior solutions, result in new interior solutions which are conceptually satisfying.

\subsection{Schwarzschild interior in the new paradigm}\label{sec:int}

\medskip
\noindent
{\bf More problems with the conventional Schwarzschild interior:}
As has been mentioned above, the constant density fluid renders the usual Schwarzschild interior unphysical. Moreover, it also causes some other conceptual problems.
Although the metric potentials of the interior and exterior line elements appear to match formally at the sharp boundary \cite{tolman}, in fact they do so only for the case $\rho=0$ as one would expect a smooth transition in the density of matter from inside to the outside of the sphere. Though several other solutions, claiming interiors to the Schwarzschild exterior solution, have also been discovered, they too suffer, more or less, from these problems.

\medskip
\noindent
{\bf Requirements for a proper interior solution:}
Isotropy is of course the primary requirement for a correct interior solution so that it can give rise to the isotropic Schwarzschild exterior solution. Further, the physical viability requires that the density of the isotropically distributed matter should not be constant. 
Rather, it should gradually diminish with $r$  finally reducing to zero, in order to achieve a smooth transition from
the field inside the sphere to the field outside the sphere.

\medskip
\noindent
{\bf The correct interior:} 
Though we do not have at present a solution of equations (\ref{eq:RicciEq}) representing such an interior, it can be approximated in the following way.
Let us divide this static interior into thin shells so that the matter distribution in any individual shell has approximately a uniform density. We have noted that the new paradigm predicts a Minkowskian geometry inside a static uniform distribution of matter, since the sum of the gravitational and inertial energies is vanishing at each point. In this view, while the contributions of gravitational and inertial energies from the individual shells would be balanced and canceled, the succeeding shells would receive extra gravitational energy from the inner shells which will show up as a non-vanishing source. This would result in a Minkowskian core surrounded by a curved envelope whose curvature increases gradually from zero to the curvature of the exterior solution at the boundary. This would also provide an excellent matching of the density and curvature of the interior and exterior at the boundary.

This chain of shells can be represented by solution (\ref{eq:Sch}) with different values of the parameter $K$ in different shells: $|K|$ gradually increasing from zero (in the core) to $2GM/c^2$ (at the boundary), where $M$ is the mass of the whole interior (which appears in the Schwarzschild exterior solution).
The value of $K$ in a particular shell will be determined by the mass of the preceding sub-sphere.

\subsection{Kerr interior}\label{sec:int-kerr}

We know that the Kerr solution (\ref{eq:kerr})
reduces to the Schwarzschild solution (\ref{eq:Sch}) for a vanishing rotation.
Hence, in the case of a vanishing rotation, the Kerr 
interior is expected to reduce to the Schwarzschild interior. That is, the Kerr interior is formed by the Schwarzschild interior plus a rotation.
This predicts that the core to the Kerr interior should be a curved, stationary, rotating spacetime solution of  equations (\ref{eq:RicciEq}) which should become flat for a vanishing rotation.
Clearly, solution (\ref{eq:int-kerr}) fulfills these requirements and hence can form the required core of the Kerr interior.

In line with the Schwarzschild interior described above, the Kerr interior will be comprised of this core of radius $2\sqrt{2}/\ell$ enveloped by the shells of depleting density of the Schwarzschild interior, but now endowed with rotation, with angular momentum per unit mass $=\alpha$, say. Obviously these shells can be similarly approximated by solution (\ref{eq:kerr}) with their corresponding values of $K$ in the Schwarzschild interior. 
The proposed Kerr-interior is vindicated by the following evidences: 

(i) it reduces to the Schwarzschild-interior in the absence of rotation, i.e., ${\cal J}=\alpha=0$, 

(ii) it is free from any singularity as is rightly expected from a smooth matter distribution, 

(iii) the matching at the boundary is smooth and satisfactory. 

\noindent
The parameter $\ell$ appearing in solution (\ref{eq:int-kerr}), can now be estimated by using the conservation of the angular momentum, which must hold. This gives
\be
{\cal J}\int \sqrt{-g}~d^3x+c\alpha~M_{\rm shell}=c\alpha M ~\Rightarrow  ~~\ell^2=\frac{64\sqrt{2}\pi c^2}{3G}\frac{1}{(M-M_{\rm shell})\alpha},
\ee
where $M_{\rm shell}$ is the total mass of the enveloping shells and $M$ is the mass of the entire interior appearing in the (exterior) Kerr solution (\ref{eq:kerr}).

\section{Einstein-Maxwell Equations in the New Paradigm}

We have seen various evidences of the presence of the gravitational and inertial (matter) fields in equations (\ref{eq:RicciEq}) without including the energy-stress tensor $T_{ik}$. What about other fields, for instance the electromagnetic field, which is another long-range interaction?

Although, there is no generally accepted unified theory of gravitation and electromagnetism, it seems reasonable to treat electromagnetism within the framework of GR. This would make a unification of the two theories at least to the extent of allowing the electromagnetic field to influence the geometry of spacetime. One would expect this unification to be done according to the  validity and manifestation of the equivalence principle in the two theories, as the equivalence principle is the defining principle of GR.  However, the usual Einstein-Maxwell equations\footnote{Equation (\ref{eq:EM}) is obtained by adding the electromagnetic energy tensor $E_{ik}$  to equation (\ref{eq:EinsteinEq}) and then considering its mixed-tensor form.} 
\be
R^i_k-\frac{1}{2} \delta^i_k R^\ell_\ell=-\frac{8\pi G}{c^4}(T^i_k + E^i_k),\label{eq:EM}
\ee
do not seem to fulfill this requirement, as is clear from the following.

\subsection{Electromagnetic field and the equivalence principle}

The implications of the equivalence principle are far deeper than obvious. The correct vision of the principle is revealed when we compare gravitation with another interaction, such as the electromagnetic one.
While the equivalence principle holds in the case of the former allowing a local cancellation (or generation) of the gravitational field by local inertial frames (owing to the fact that the ratio of the gravitational and inertial mass is strictly unity for all matter), this is not so in the case of the latter
(where the ratio of electric charge to mass varies from particle to particle). In the latter case the  field is described within a constant and immutable arena of spacetime. Whereas in the former case, the equivalence principle makes it possible to represent gravitation as a metric phenomena, wherein  the spacetime itself becomes a dynamical object. This suggests that gravity is fundamentally different from the electromagnetic  (or the weak and strong) forces and we should expect this difference to be reflected in the theory. However, equations (\ref{eq:EM}) fail to express any such difference and the sources of the two interactions, viz., the matter tensor $T^i_k$ and the electromagnetic energy tensor $E^i_k$, are treated by the equation on the same footing and couple to the geometry in the same way.

On the other hand, the new paradigm, in accordance with the equivalence principle (which renders gravitation as a geometric phenomenon), manifests the gravitational as well as the inertial fields through the geometry (consistent with the local equivalence of gravitation and inertia) and {\it not through} $T^i_k$, thereby constituting equations (\ref{eq:RicciEq}) as the field equations of pure gravitation (plus inertia).

As this kind of geometrization would not be possible in the case of the electromagnetic forces, the electromagnetic properties of matter would need a tensor representation  and equations (\ref{eq:EinsteinEq}) [{\it not equations (\ref{eq:EM})}] would represent, in the new paradigm, the field equations of gravitation (plus inertia) plus electromagnetism with $T_{ik}$ serving as the electromagnetic energy tensor, showing the expected difference between the two interactions in the context of the equivalence principle. This is also corroborated by the fact that we do need the energy tensor $E^i_k$ in order to obtain, for instance, the Reissner-Nordstrom and the Kerr-Newman solutions, which should be considered as a reflection of the fact that no equivalence principle (to the degree of the one for the gravitational force) exists for the electromagnetic forces.

We are hence led to believe that the above-mentioned `full-geometrization of matter' is, in fact, a correct realization of the equivalence principle, implying a novel way of manifestation of the inertial plus gravitational properties of matter thoroughly through the geometry (and not through the matter tensor $T^i_k$), in contrast with the electromagnetic properties of matter, which must have the usual representation through the electromagnetic energy tensor $E^i_k$.
Hence, identifying $T_{ik}$ with the  electromagnetic energy tensor $E_{ik}$  in equations (\ref{eq:EinsteinEq}) (in which case  $R=0$, as $g^{ik} E_{ik}=0$ identically), the field equations (in mixed-tensor form) yield
\be
R^i_k=-\frac{8\pi G}{c^4}E^i_k,\label{eq:e-m-corr}
\ee
which would represent, in the new paradigm, the field equations of gravitation (plus inertia) plus electromagnetism
 in the real Universe with matter.  

As the well-known solutions of equations (\ref{eq:e-m-corr}) -  the Reissner-Nordstrom and the Kerr-Newman solutions - represent field {\it outside} the charged matter source,
it is generally believed that equations (\ref{eq:e-m-corr}) can support only those meaningful solutions which describe the field outside some charged matter.
But what if equations (\ref{eq:e-m-corr}) provide a cosmological solution, which is not expected to contain any `outside' where the charge-carrier matter can exist?  Would it then not mean that the matter is already present in the spacetime, as the presence of charge can only be expected in the presence of the charge-carrier matter? So, let us try to electrify the Milne model (which represents the real Universe in the new paradigm) through equations (\ref{eq:e-m-corr}). If equations (\ref{eq:e-m-corr}) can support an electromagnetic field for this model, that would be a powerful evidence of the correctness of the new paradigm!

\subsection{Equations of electrodynamics in the presence of gravitation}

The Maxwell equations, in the presence of a gravitational field, are conventionally given by the following  pair of equations
\be\label{eq:Max}
\left.\begin{aligned}
\frac{\partial F_{ik}}{\partial x^\ell}+\frac{\partial F_{k \ell}}{\partial x^i}+\frac{\partial F_{\ell i}}{\partial x^k}=0\\
F^{ik}_{~~;k}=\frac{1}{\sqrt{-g}}\frac{\partial}{\partial x^k}(\sqrt{-g}F^{ik})=\xi j^i
 \end{aligned}
 \right\},
\ee
where $F_{ik}$ is the skew-symmetric electromagnetic field tensor, $g={\rm det}((g_{ik}))$ is the determinant of the metric tensor, $\xi$ is a constant, $j^i$ is the four-current density vector and a semicolon followed by a lower index denotes partial derivative with respect to the corresponding variable.
As the existence of charge is intimately related with the existence of the charge-carrier matter and since the new paradigm claims the inherent presence of matter in the geometry, it would be reasonable to expect the charge also to appear through the geometry. This view is supported by the Reissner-Nordstrom and Kerr-Newman solutions wherein  the source charge does appear through the geometry, though the charge density is considered vanishing in the second pair of equations (\ref{eq:Max}), which then reduce to 
\be\label{eq:Max-corr}
\left.\begin{aligned}
\frac{\partial F_{ik}}{\partial x^\ell}+\frac{\partial F_{k \ell}}{\partial x^i}+\frac{\partial F_{\ell i}}{\partial x^k}=0\\
\frac{\partial}{\partial x^k}(\sqrt{-g}F^{ik})=0
 \end{aligned}
 \right\}.
\ee
Equations (\ref{eq:Max-corr}) are usually believed to represent the electromagnetic field in vacuum (very much in the same vein as equations $R_{ik}=0$ are considered to represent the gravitational field in vacuum in the standard paradigm).  Nevertheless,  equations (\ref{eq:Max-corr}) are supported not only by the Reissner-Nordstrom and Kerr-Newman solutions representing the field outside the charged matter, but they are also supported by the cosmological solutions - the so-called `electrovac universes' \cite{electrovac}, which though have not been paid much attention and have remained just a mathematical curiosity.

We believe that this is not another big coincidence and the situation somewhat resembles the earlier discussed case where  the correct Machian strategy makes the energy-stress tensor $T_{ik}$ redundant in Einstein's equations (\ref{eq:EinsteinEq}). Here, equations (\ref{eq:Max-corr}) represent electromagnetic field in the very presence of matter and  the four-current density vector  $j^i$ is rendered redundant. This may be considered as a  `semi-geometrization' of the electromagnetic field in the new paradigm, as should be expected from a geometric theory of gravitation plus electromagnetism.

\subsection{Electromagnetic field in Milne's Universe}

To make the calculations simple, let us consider the Milne model in its manifestly-Minkowskian form. Let us also consider the geometric units ($c=G=1$) henceforth to give some respite to already heavily symbol-laden theory. 
In order to support the electromagnetic field, let us consider a simple generalization of the Minkowski metric:
\be
ds^2= dt^2-dx^2-dy^2-(1+a_t t +a_x x+a_y y)^{2m} dz^2,\label{eq:milneEM}
\ee
where the constant $m$ is different from 0 and 1 (in which cases the metric becomes Minkowskian) and the constants $a_i$ vanish when the electromagnetic field switches off. These constants are to be evaluated by solving equations  (\ref{eq:e-m-corr}) wherein the tensor $E^i_k$  is given by
\be
E^i_k=-F^{ij}F_{kj}+\frac{1}{4}\delta^i_k F_{\ell h}F^{\ell h}\label{eq:e-m}
\ee
and the electromagnetic field tensor $F_{ik}$ is to satisfy the Maxwell equations (\ref{eq:Max-corr}). 
The Ricci scalar $R$ for metric (\ref{eq:milneEM}) comes out as
\be
R=\frac{2m(a_t^2-a_x^2-a_y^2)(m-1)}{(1+a_t t+a_x x+a_y y)^2}\label{eq:R}.
\ee
As equations (\ref{eq:e-m-corr}) imply that $R=0$, this when plugged in equation (\ref{eq:R}),  supplies $a_t=\pm\sqrt{a_x^2+a_y^2}$ as the only non-trivial solution.    Let us assume $a_y=0$ to simplify the matter further, which gives $a_t=\pm a_x$. In order to retain $g_{zz}$ negative definite for all possible values of $m$ in (\ref{eq:milneEM}), we consider $a_t= a_x$ ($= a$, say) with $a>0$ reducing (\ref{eq:milneEM}) to
\be
ds^2= dt^2-dx^2-dy^2-(1+a x + a t)^{2m} dz^2.\label{eq:milneEMfin}
\ee
For this line element, the non-vanishing components of $R^i_k$ are obtained as
\be
R^t_t=R^t_x=-R^x_x=-\frac{a^2m(1-m)}{(1+a x + a t)^2}.\label{eq:Es}
\ee
Since the time-time component of $E^i_k$ given by (\ref{eq:e-m}), i.e.
\be
E^t_t=\frac{1}{2}\big[(g^{xx}g^{yy}F_{xy}^2+g^{yy}g^{zz}F_{yz}^2+g^{zz}g^{xx}F_{zx}^2)-g^{tt}(g^{xx}F_{xt}^2+g^{yy}F_{yt}^2+g^{zz}F_{zt}^2)\big],
\ee
turns out to be a positive definite quantity (for a meaningful non-vanishing electromagnetic field), the component $R^t_t$ must be negative in order to satisfy equation  (\ref{eq:e-m-corr}). This condition determines $m$ as $0<m<1$.

We find the following two solutions of equations (\ref{eq:e-m-corr}), (\ref{eq:e-m}) and (\ref{eq:Es}), which also satisfy (\ref{eq:Max-corr}):

\begin{equation}\label{eq:Fs1}
 \left.\begin{aligned}
F_{xt}=F_{xz}=F_{zt}=F_{yz}=0,\\
F_{xy}=-F_{yt}=\pm\frac{a\sqrt{m(1-m)}}{\sqrt{8\pi}(1+a x+ a t)}
       \end{aligned}
 \right\},
\end{equation}\\
\begin{equation}\label{eq:Fs1}
 \left.\begin{aligned}
F_{xt}=F_{xy}=F_{yt}=F_{yz}=0,\\
F_{zt}=-F_{xz}=\pm\frac{a\sqrt{m(1-m)}}{\sqrt{8\pi}(1+a x+ a t)^{1-m}}
       \end{aligned}
 \right\}.
\end{equation}
These solutions provide different possible distributions of the electromagnetic field supported by the Milne Universe.
Since at least one of the components $F_{xt}$, $F_{yt}$, $F_{zt}$ is non-zero in these solutions, they guarantee the presence of charge (density), and hence matter, in the spacetime  geometry.

\section{Conclusion}

Mach was primarily responsible for  denying Newton's unobservable absolute space in favor of the observable quantities, which shaped special relativity as well as Heinsenberg's and others' work on quantum mechanics.
His ideas have enormous importance in the historical development of GR which was at least conceived in the spirit of realizing his philosophy. Nevertheless the theory could not fulfill this expectation.

The present paper develops a deeper insight into Mach's principle revealing that
 the spacetime has no independent existence without a background of matter fields which are built-in ingredients of the geometry of the canonical field equations $R_{ik}=0$.
 This constitutes a new paradigm in GR wherein Mach's principle finds its clearest expression, as the spacetime structures is determined by the net contribution from the material and the gravitational fields.
The new paradigm appears more comprehensive and
insightful than the conventional one and grants the spacetime an existence with physical
qualities of its own. By approaching the very foundation of our ideas of spacetime, the new paradigm
gets more and more to the status envisioned initially by Einstein.

Besides explaining the observations at all scales, the new paradigm provides an appealing first-principle approach towards answering various seemingly insoluble puzzles and inexplicable coincidences confronting GR and cosmology.
New interior solutions, forming the Schwarzschild interior and the Kerr interior, are proposed in the new paradigm, which are conceptually satisfying.

A new solution of $R_{ik}=0$, predicted by the new paradigm, is discovered. The solution is inhomogeneous and axisymmetric whose curvature is sourced by a dimensional parameter sustaining the energy density. 
A new (so-called) `electro-vac' solution is also discovered which corroborates the new paradigm.


\begin{thebibliography}{}




\bibitem{progress}  W.-T. Ni, Empirical Foundations of the Relativistic Gravity, {\it Int. J. Mod. Phys. D} {\bf 14} (2005) 901-921;
C. M. Will, The confrontation between general relativity and experiment, {\it Living Rev. Rel.} {\bf  9} (2006) 3;
S. G. Turyshev, Experimental tests of general relativity: recent progress and future directions, {\it Phys.-Usp.} {\bf 52} (2009) 1. 

\bibitem{centennial} L. Iorio, Editorial for the special issue 100 Years of chronogeometrodynamics: The status of the Einstein's theory of gravitation in its centennial year, {Universe} {\bf 1} (2015) 38-81;
 W.-T. Ni, {\it One Hundred Years of General Relativity. From Genesis and Empirical Foundations to Gravitational Waves, Cosmology and Quantum Gravity}, (In 2 Volumes), (World Scientific, 2015).

\bibitem{Mach} E. Mach, {\it The Science of Mechanics}, (Open Court, La Salle, 1960).

\bibitem{Mach-many} J. Barbour and H. Pfister (Editors), {\it Mach's Principle: From Newton's Bucket to Quantum Gravity}, (Birkhäuser, Boston, 1995);

\bibitem{Godel} K. G$\ddot{o}$del, An example of a new type of cosmological solution of Einstein's field equations of gravitation, {\it Rev. Mod. Phys.} {\bf 21} (1949) 447-450.

\bibitem{Rindler} W. Rindler, The Lense-Thirring effect exposed as anti-Machian, {\it Phys. Lett. A} {\bf 187} (1994) 236-238;
W. Rindler, The case against space dragging, {\it Phys. Lett. A} {\bf 233} (1997) 25-29;
 H. Lichtenegger and B. Mashhoon, Mach's Principle, in {\it The Measurement of Gravitomagnetism: A Challenging Enterprise}, (Ed. L. Iorio) (Nova Science, New York, 2007) 13-25.

\bibitem{Bondi-Samuel} H. Bondi and J. Samuel, The Lense-Thirring effect and Mach's principle, {\it Phys. Lett. A} {\bf 228} (1997) 121-126;
A. R. Prasanna, Inertial frame dragging and Mach’s principle in General Relativity, {\it Class. Quant. Grav.} {\bf 14} (1996) 227-236.

\bibitem{OS} I. Ozsv$\acute{a}$th and E. Sch$\ddot{u}$cking,  An anti-Mach metric, in {\it Recent Developments in
General Relativity} (Ed. L. Witten), (Pergamon Press, Oxford, 1962) 339-350.

\bibitem{Taub} A. Taub, Empty Space-Times Admitting a Three Parameter Group of Motions, {\it Ann. of Math.} {\bf 53}  (1951) 472-490; 
EmptySpace Generalization of the Schwarzschild Metric, {\it J. Math. Phys.} {\bf 4} (1963) 915-923.

\bibitem{Paper1} R. G. Vishwakarma, A new solution of Einstein's vacuum field equations',  {\it Pramana - J. Phys.} DOI: 10.1007/s12043-015-0946-3  [arXiv: 1409.3758] (2015). 

\bibitem{Misner} C. Misner, The Flatter Regions of Newman, Unti, and Tamburino's Generalized
Schwarzschild Space, {\it J. Math. Phys.} {\bf 4} (1963) 924-937.

\bibitem{curious} V. V. Narlikar and K. R. Karmarkar, On a curious solution of relativistic field equations, {\it Curr. Sci.}  {\bf 15} (1946) 69. 

\bibitem{Einstein} A. Einstein,  {\it Relativity: the Special and the General Theory}, (1955).

\bibitem{vishwaApSS2} R. G. Vishwakarma, On the relativistic formulation of matter, {\it Astrophys. Space Sci.} {\bf 340} (2012) 373-379.

\bibitem{Vishwa_Front.Phys.} R. G. Vishwakarma, Mysteries of $R^{ik}$ = 0: A novel paradigm in
Einstein’s theory of gravitation, {\it Front. Phys.} {\bf 9} (2014) 98–112.

\bibitem{narlikar} J. V. Narlikar, {\it An Introduction to Cosmology}, (Cambridge University Press, 2002).

\bibitem{gravitoEM} R. Maartens and  B. A. Bassett, Gravito-electromagnetism, {\it Class. Quant. Grav.} {\bf 15} (1998) 705-717.

\bibitem{Bahram1} D. Bini, C. Cherubini, C. Chicone and B. Mashhoon, Gravitational induction, {\it Class. Quant. Grav.} {\bf 25} (2008) 225014.

\bibitem{Sciama} D. W. Sciama, On the origin of inertia, {\it MNRAS} {\bf 113} (1953) 34-42.

\bibitem{Hawking-Milodinow} S. Hawking and L. Milodinow, {\it The Grand Design}, (Bantom Books, New York,  2010).

\bibitem{Schmid} C. Schmid,  Cosmological gravitomagnetism and Mach's principle, {\it Phys. Rev. D} {\bf 74} (2006) 044031; 
C. Schmid, Mach's principle: Exact frame-dragging via gravitomagnetism in perturbed Friedmann-Robertson-Walker universes with $k$=($\pm$1, 0), {\it Phys. Rev. D} {\bf 79} (2009) 064007.

\bibitem{testGravitoMag}
L. Iorio, Is it possible to improve the present LAGEOS Lense-Thirring experiment?, {\it Class. Quant. Grav.} {\bf 19} (2002) 5473-5480;
I. Ciufolini, Frame dragging and Lense-Thirring effect, {\it Gen. Rel. Grav.} {\bf 36} (2004) 2257-2270;
L. Iorio, Evidence of the gravitomagnetic field of Mars, {\it Class. Quant. Grav.} {\bf 23} (2006) 5451-5454;
L. Iorio,First evidence of the general relativistic gravitomagnetic field of the Sun and new constraints on a Yukawa-like fifth force, {\it Planetary Space Sci.} {\bf 55} (2007) 1290-1298;
C. W. F. Everitt,  et al., Gravity Probe B: Final results of a space experiment to test General Relativity, {\it Phys. Rev. Lett.} {\bf 106} (2011) 221101;
L. Iorio, H. I. M. Lichtenegger, M. L. Ruggiero and C. Corda, Phenomenology of the Lense-Thirring effect in the Solar System, {\it Astrophys. Space Sci.} {\bf 331} (2011) 351-395;
L. Iorio, Constraining the angular momentum of the Sun with planetary orbital motions and general relativity, {\it Solar Phys.} {\bf 281} (2012) 815-826;
A. Paolozzi and I. Ciufolini, LARES successfully launched in orbit: Satellite and mission description, {\it Acta Astronautica} {\bf 91} (2013)  313-321; 
L. Iorio, et al., Novel considerations about the error budget of the LAGEOS-based tests of frame-dragging with GRACE geopotential models, {\it Acta Astronautica} {\bf 91} (2013) 141-148; 
G. Renzetti, First results from LARES: An analysis, {\it New Astronomy} {\bf 23-24} (2013) 63-66; 
G. Renzetti, Some reflections on the Lageos frame-dragging experiment in view of recent data analyses, {\it New Astronomy} {\bf 29} (2014) 25-27; 
G. Renzetti, On Monte Carlo simulations of the LAser RElativity Satellite experiment, {\it Acta Astronautica} {\bf 113} (2015) 164-168. 
\bibitem{vishwa_milne} R. G. Vishwakarma, A curious explanation of some cosmological phenomena, {\it Phys. Scripta} {\bf 5} (2013) 055901. 

\bibitem{Bondi} H. Bondi,  {\it Cosmology}, (Cambridge University Press, second edition,  1968).

\bibitem{Noether} E. Noether, Invariant variation problems, {\it Transp.Theory Statist. Phys.} {\bf 1} (1971) 186; 
Yu. V. Baryshev,  Energy-momentum of the gravitational field: crucial point for gravitation physics and cosmology, {\it Practical Cosmology} {\bf 1} (2008) 276-286.

\bibitem{NewtonianGrav} P. Peebles, {\it The large-scale structure of the universe}, (Princeton, NJ: Princeton
University Press, 1980); J. V. Narlikar and G. R. Burbidge, An empirical approach to cosmology, {\it Astrophys. Space Sci.} {\bf 74} (1981) 111.

\bibitem{Ishak} M. Ishak, W. 5, J.  Dossett, J.  Moldenhauer and C.  Allison, A new independent limit on the cosmological constant/dark energy from the relativistic bending of light by galaxies and clusters of galaxies,
{\it MNRAS} {\bf 388} (2008) 1279.

\bibitem{tolman}  R. C. Tolman, {\it Relativity, Thermodynamics and
Cosmology}, (Oxford University Press, 1934).

\bibitem{electrovac} M. Misra, Electrovac universes, {\it Proc. Nat. Inst. Sci.} {\bf 28A} (1962) 105-119; 
R. Tiwari and M. Misra, Axially symmetric conformastat electrovac universes, {\it Proc. Nat. Inst. Sci.} {\bf 28A} (1962) 857-862;
L. Radhakrishna, Some exact non-static cylindrically symmetric electrovac universes, {\it Proc. Nat. Inst. Sci.} {\bf 29A} (1963) 588-595;
K. P. Singh and Abdussattar, Some cylindrically symmetric electrovac universes, {\it Proc. Ind. Nat. Sci. Acad.} {\bf 43} (1977) 465-469.



\end{thebibliography}
\end{document}